\documentclass[aps,pop,twocolumn,superscriptaddress,letterpaper]{revtex4}
\usepackage{mathrsfs}
\usepackage{amsmath}
\usepackage{bm}
\usepackage{graphicx}
\usepackage{hyperref}
\usepackage{CJK}

\begin{document}
\begin{CJK*}{GB}{gbsn}


\title{Constant residual electrostatic electron plasma mode in Vlasov-Ampere system}
\author{Hua-sheng XIE}
\email[]{Electronic mail: huashengxie@gmail.com}
\affiliation{Institute for Fusion Theory and Simulation, Department
of Physics, Zhejiang University, Hangzhou, 310027, People's Republic
of China}
\date{\today}

\begin{abstract}
In a collisionless Vlasov-Poisson (V-P) electron plasma system, two
types of modes for electric field perturbation exist: the
exponentially Landau damped electron plasma waves and the
initial-value sensitive ballistic modes. Here, the V-P system is
modified slightly to a Vlasov-Ampere (V-A) system. A new constant
residual mode is revealed. Mathematically, this mode comes from the
Laplace transform of an initial electric field perturbation, and
physically represents that an initial perturbation (e.g., external
electric field perturbation) would not be damped away. Thus, this
residual mode is more difficult to be damped than the ballistic
mode. [Physics of Plasmas 20, 112108 (2013); doi: 10.1063/1.4831761]
\end{abstract}


\maketitle

\section{Introduction}\label{sec:intro}
For the evolution of linear electron plasma wave in an unmagnetized
plasma, the one-dimensional linearized Vlasov-Poisson (V-P) system
is
\begin{subequations}\label{eq:es1d}
\begin{eqnarray}
{\partial_t\delta f}&=&-ikv\delta f+\delta E{\partial_v{f_0}}, \label{eq:vlasov} \\
ik\delta E &=&-\int{\delta fdv}, \label{eq:poisson}
\end{eqnarray}
\end{subequations}
where time $t$ and space $x$ have been normalized by the inverse
plasma frequency $\omega_p^{-1}=1/\sqrt{n_0e^2/\epsilon_0m}$ and the
Debye length $\lambda_D=\sqrt{\epsilon_0\kappa_BT/n_0e^2}$,
respectively, and $\kappa_B$ is Boltzmann's constant. That is,
$\omega_p=1$ and $\lambda_D=1$. The perturbations have the harmonic
dependence  $e^{ikx}$, and $f_0=f_M=\exp(-v^2/2)/\sqrt{2\pi}$, where
$v$ has been normalized by the electron thermal speed, is the
Maxwellian equilibrium distribution.

In most cases, the plasma frequency is sufficiently high compared
with the electron collision frequency $\nu$. The collision can be
neglected for time scale $t\ll \nu^{-1}$. The typical characteristic
nonlinear time scale $t_{NL}$ is the bounce time
$t_B=\omega_B^{-1}=(k\delta E)^{-1}$ of trapped particles. Due to
$\delta E\ll 1$,
$\omega^{-1}\simeq\omega_{p}^{-1}=1\ll\omega_B^{-1}\simeq t_{NL}$,
we can also ignore the nonlinear effect, because we are mainly
interested in the linear time scale in this paper, i.e., $t\ll
t_{NL}$. We have also assumed the immobile ion in Eq.(\ref{eq:es1d})
because $m_e/m_i\leq1/1836\ll1$.

As an initial value problem, for arbitrary initial $\delta f(v)$,
the problem can be described by a superposition of a complete set of
eigen solutions, as shown by van Kampen\cite{Kampen1955} and
Case\cite{Case1959}. It is also well known that this system can
support exponentially Landau damped mode\cite{Landau1946}, which is
the time asymptotic behavior. This asymptotic behavior can also be described by
dispersion relation\cite{Landau1946,Jackson1960}. For convenience, we designate the dispersion
relation solution portion as the Landau part and the initial-value
sensitive [referring to the moments of the perturbations, as shown
below, or Eqs.(\ref{eq:EktVAlim}) and (\ref{eq:EktVPlim})] portion
in Case-van Kampen (CvK) mode as the ballistic part, as the latter
part is usually closely related to the $-ikv\delta f$ term in
Eq.(\ref{eq:es1d}). For $f_0=f_M$, CvK eigen solutions are undamped
in the sense that no solution of the linear Vlasov-Poisson equation
with $\Im(\omega)<0$  exist for the harmonic perturbation
$\sim\exp(-i\omega t)$, where $\omega$ is the mode frequency.
However, because of phase mixing, the (moments with respect to
velocity of the) perturbations are still damped when all such modes
are considered, e.g., considering $\delta E(t)$ instead of $\delta
f(v,t)$, which can yield Landau's solution for some initial $\delta
f(v)$ (especially, entire function\cite{Landau1946, Stix1992}, which
is holomorphic over the whole complex plane). That is, the ballistic
modes (if not specified, here and hereafter, we are referring to the
moments of these modes) usually decay faster than Landau damping. A
comprehensive description of these modes and their relations in this
V-P system can be found in Ref.\cite{Jackson1960}.

For example, for the initial perturbation $\delta
f(v,0)=A_0\exp[-(v-u_a)^2/u_b^2]$, where $|u_a|\gg(|u_b|,
|\omega_r/k|)$ with $\omega_r$ real frequency of the least damping
Landau solution, the ballistic mode decays like
$\exp(-iku_at-k^2u_b^2t^2)$ \cite{Chen1987}. On the contrary, for
example, a slow (relative to that of Landau) algebraic decay mode
\begin{equation}\label{eq:timeasym}
\propto \frac1t\sin(ku_bt)\exp(-iku_at),
\end{equation}
can be excited by the initial perturbation \cite{Chen1987}
\begin{equation}\label{eq:df0}
\delta f(v,0)=\left\{
\begin{array}{ll}
A_0, & \mbox{for $|v-u_a|\leq u_b$},\\
0, & \mbox{otherwise},
\end{array}
\right.
\end{equation}
which is not holomorphic, and thus not an entire function.

If we neglect the term proportional to the electric field in
Eq.(\ref{eq:es1d}), the solution Eq.(\ref{eq:timeasym}) will be
straightforward and precise. However, the difference is that, for
Eq.(\ref{eq:timeasym}), the electric field is kept, which holds for
very large $t$, i.e., $t\to \infty$.

For the linear collisionless initial value problem of the V-P
system, the above pictures are complete.

In this paper, we find that if we change the equations slightly to
Vlasov-Ampere (V-A) equations, the dominant mode will be a new
constant residual mode, and not the Landau and ballistic modes.

\section{Vlasov-Ampere system}\label{sec:VA}
From the charge continuity equation
${\partial_t\rho}+{\partial_xJ}=0$, where $\rho=\int{\delta fdv}$
and $J=\int{v\delta fdv}$, the V-A and V-P systems are equivalent (the proof is straightforward) if
the Poisson's equation is used to supplement the initial condition
in the V-A system.

The reasons below fueled our interest to obtain a more complete
description of the linearized V-A system, where the Poisson's
equation Eq.(\ref{eq:poisson}) in the V-P system has been replaced by
Ampere's law Eq.(\ref{eq:ampere}) and the whole system of
Eq.(\ref{eq:es1d}) is changed to
\begin{subequations}\label{eq:es1dva}
\begin{eqnarray}
{\partial_t\delta f}&=&-ikv\delta f+\delta E{\partial_v{f_0}}, \label{eq:vavlasov} \\
{\partial_t\delta E}&=&\int{v\delta fdv}, \label{eq:ampere}
\end{eqnarray}
\end{subequations}
where the normalization and all other definitions are unchanged.

First, in the work of Horne and Freeman \cite{Horne2001}, they
needed to use the V-P equations for the first few steps of the
numerical integration before continuing with the V-A equations.
Without the Poisson start, recovering the Landau damping solutions
is difficult. How do we explain this finding?

Second, in a toroidal system, a well-known residual mode called the
Rosenbluth-Hinton residual zonal
flow\cite{Rosenbluth1998,Hinton1999}, was found. The authors showed
that poloidal flows driven by ion-temperature gradient (ITG)
turbulence will not damp to zero by linear collisionless processes,
and the final poloidal velocity $u_p\neq0$\cite{Rosenbluth1998}.
Usually, physics in complicated systems can be understood using
simple system. For example, the phase mixing in velocity space can
also be found in a real space non-uniform system (see
e.g., \cite{Hasegawa1974,Qiu2011}), and rich physics
are associated. Thus, phase mixing is a useful concept for
understanding continuum damping, from Alfv\'en wave\cite{Hasegawa1974} to geodesic acoustic mode\cite{Qiu2011}.
Can a constant residual mode be found in a simple system?

\subsection{Initial value solutions}

Before to discuss the new constant residual mode, one can refer to
the review of the derivations and the verification of the Landau and
ballistic solutions in V-P system in the Appendix \ref{sec:vp},
which will be used for comparison.

Similar to the V-P system, the dispersion relation of plasma waves
for the V-A system can be derived as
\begin{equation}\label{eq:drva}
    D_{\mbox{VA}}(\omega,k) \equiv \omega+\int_C{\frac{v{\partial
    f_0}/{\partial v}}{\omega-kv}dv}=0.
\end{equation}

Eqs.(\ref{eq:drva}) and (\ref{eq:drvp}) are equivalent, so they should
both yield the normal modes given by the dispersion relation, which
(the asymptotic solution) is independent of the initial condition.
That is, we should also be able to obtain the Landau damped
solutions from the V-A equations without using the Poisson start.
However, as mentioned, existing simulations indicate that this
process is difficult.

Applying Laplace transform ($L_p$) in time and Fourier transform
($F_r$) in space to the V-A system, we can obtain
\begin{eqnarray}\label{eq:EktVA}
    \delta E(t,k) &=& L_p^{-1}\delta \hat{E}(\omega,k) \\
        &=& \int_{C_\omega}{\frac{e^{-i\omega t}d\omega}{2\pi}}\Big [\int{dv\frac{v\delta f_k(0)}{(\omega-kv)D(\omega,k)}}-\frac{\delta E_k(0)}{D(\omega,k)} \Big], \nonumber
\end{eqnarray}

We are interested in asymptotic behavior, so only the $\omega=kv$
pole and the maximum-$\Im\omega_m$ normal mode are kept. From
Eq.(\ref{eq:EktVA}) we have,
\begin{eqnarray}\label{eq:EktVAlim}
    \lim_{t \to \infty}\delta E &=& {\frac{1}{2\pi}}\int_{v_L}{dv\Big [\underbrace{\frac{ve^{-ikvt}\delta f_k(0)}{D_{\mbox{VA}}(kv,k)}}_{\mbox{ballistic part}} +\underbrace{\frac{ve^{-i\omega_mt}}{(\omega_m-kv)\partial_{\omega_m}
    D}}_{\mbox{Landau part}}
    \Big]} \nonumber \\
& & -\underbrace{\int_{C_\omega}{\frac{e^{-i\omega t}d\omega}{2\pi}}\frac{\delta E_k(0)}{D_{\mbox{VA}}(\omega,k)}}_{\mbox{initial $\delta E$ part}}.
\end{eqnarray}

Noting the relation $kD_{\mbox{VA}}=\omega D_{\mbox{VP}}$, the main
difference between Eqs.(\ref{eq:EktVAlim}) and (\ref{eq:EktVPlim}) is in
the initial $\delta E$ or $\delta E(0)$, which is a constant. To
determine the typical asymptotic behavior, we let
$D_{\mbox{VA}}\approx \omega$ in the $\delta E$ part of the integral
and obtain the residual mode
\begin{equation}\label{eq:EktVAlim2}
\lim_{t \to \infty}[\delta E_{\mbox{VA}}(t) -\delta
E_{\mbox{VP}}(t)+\delta E_{\mbox{Poisson}}(0)] \propto \delta E(0).
\end{equation}
Eq.(8) indicates that for almost all kinds of initial perturbations $\delta f(v,0)$,
the electric field perturbation $\delta E$ will
not be damped away, except that when the initial $\delta E$
satisfies the Poisson's
equation Eq.(\ref{eq:poisson}).
Therefore, in the V-A system, the dominant mode would be this constant residual mode,
and Landau damping will not be evident. Similar to the ballistic
mode, this mode is also from the initial perturbation. We shall
designate it as the E mode, to distinguish it from ballistic mode.

This result indicates that the linear residual mode can also be
found in simple systems. The issue on why we can hardly find Landau
damping in V-A system is also answered.

However, in contrast to the zero poloidal and toriodal mode numbers
for residual zonal flow [$m=n=0$ with perturbation quantities factor
$e^{i(m\theta-n\phi)}$], the wave vector $k$ for E mode is not
necessarily zero.

\begin{figure}
  \includegraphics[width=8.5cm]{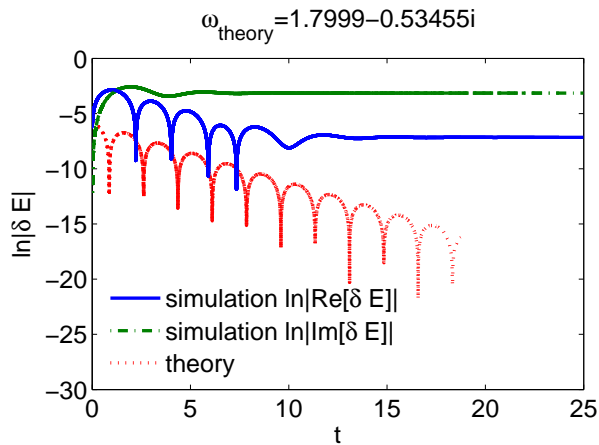}\\
  \caption{V-A simulation of the residual E mode, which remains as a non-zero constant at large $t$.
  The theoretical value $\omega_{\rm theory}=1.7999-0.53455i$ and corresponding red dot line are the
  Landau damping dispersion relation prediction, which is solved from Eq.(\ref{eq:drva}).}\label{fig:va_Emode}
\end{figure}

Now, we verify the above calculation using simulation.
Eq.(\ref{eq:es1dva}) is solved numerically as an initial value
problem. The simulation scheme and the initial $\delta f$
perturbation are similar to that in the Appendix \ref{sec:vp}. Fig.\
\ref{fig:va_Emode} shows the residual E mode, where $\delta
E(0)=0.02$.

\begin{figure}
  \includegraphics[width=8.5cm]{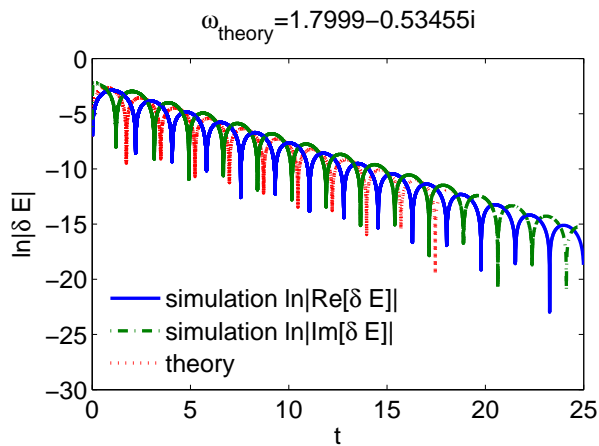}\\
  \caption{V-A simulation of Landau damping using Poisson start. At
  this condition, the V-A system is completely equivalent to the V-P system and produces Landau damping as expected.}\label{fig:va_landau}
\end{figure}

The results for a V-A simulation of Landau damping using a Poisson
start is given in Fig.\ \ref{fig:va_landau}. The parameters are the
same as those in Fig.\ \ref{fig:va_Emode}, except that a V-P run is
added in the initial $i_t$ time step(s). In this linear simulation,
$i_t=1$ is sufficient. For nonlinear simulations, one may need more
time steps \cite{Horne2001}. In Fig.\ \ref{fig:va_landau}, we can
see that both the real and imaginary parts of $\delta E$ match the
Landau result.

In the V-A simulation, the residual E mode is dominant and the
initial perturbation Eq.(\ref{eq:df0}) also cannot prevent it (not
shown here).

Compared with the V-P system, the free parameter $\delta E(0)$ in
the V-A system change the picture completely. Mathematically, the
residual mode comes from the Laplace transform of initial electric
field perturbation and physically represents that an initial mode
(e.g., external electric field perturbation) would not be damped
away. Here, we solved the initial value problem. Experimentally, we
usually address the boundary value problem, which means the electric
field will not be zero but remains a constant at large distance. The
solution for this boundary value problem can be found in
Landau\cite{Landau1946}.

\subsection{Eigenmodes in the V-A system}
Other phenomena are related to Landau damping. For example, the
Landau damped normal mode is not an eigenmode in the V-P system,
although a growing normal mode can be an eigenmode (see, e.g., Ref.\
\onlinecite{Bratanov2012}). The eigenmode problem with collision
term and the connections between collisionless case have been
studied by several
authors\cite{Ng1999,Ng2004,Bratanov2013,Bratanov2012,Hilscher2013}.
Bratanov {\it et al.}\cite{Bratanov2012,Bratanov2013} reported new
numerical investigations regarding the connection between the CvK
eigenmode and the Landau normal mode. Spectral density accumulation
occurs around the real frequency of the Landau-damped mode.

We rewrite the governing equations into the matrix form
$M\cdot\vec{F}=\omega\vec{F}$, with $\partial_t=-i\omega$. The
eigenvector $\vec{F}$ for the V-P system is
$\vec{F}=\{F_j\}=\{\delta f(v_j)\}$, where $v_j=v_{\min}+(j-1)\Delta
v$ and $j=1,2,3,...,N_v+1$. Bratanov\cite{Bratanov2012} verified
numerically that although all the (CvK) eigenvalue solutions of
$\vec{F}$ are undamped, their integral, which is related to $\delta
E$, can yield Landau damping because of phase mixing.

A question then arises: what if $\delta E$ (the moment) is also
contained in the eigenvector $\vec{F}$? Will it give the Landau
solution directly, so that the Landau damped mode can also be an
eigenmode? The V-A system can combine $\delta E$ into $\vec{F}$
directly. The eigenvector of the V-A system then becomes
\begin{equation}\label{eq:vecFva}
    \vec{F}=\{F_j\}=\left\{
            \begin{array}{ll}
            \delta f(v_j), & j=1,2,3,\cdots,N_v+1,\\
            \delta E, & j=N_v+2.
            \end{array}
        \right.
\end{equation}
Using Eqs.(\ref{eq:es1d}) and (\ref{eq:es1dva}), we can easily find the
elements of the eigenvalue matrices $M$ for the V-P and V-A systems,
respectively.

\begin{figure}
  \includegraphics[width=9.0cm]{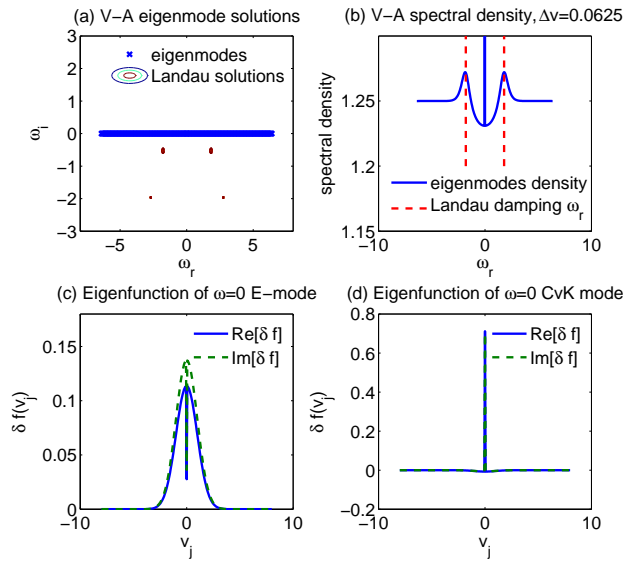}\\
  \caption{Eigenmodes in the V-A system. (a)
  compared with the Landau damping normal modes (dispersion relation solutions), (b) the spectral density,
  (c) and (d) the corresponding eigen functions for $\omega=0$ E mode and continuum CvK mode.}\label{fig:eigenmode}
\end{figure}

The solutions are shown in Fig.\ \ref{fig:eigenmode}. No eigenmodes
and the normal modes are identical in panel (a). That is, our result
also indicates that the Landau damped mode in the V-A system is not
an eigenmode.

The only difference between the V-A and V-P systems is in the
spectral density around $\omega_r=0$: an extra accumulating point
exists at $\omega_r=0$ of the V-A system [see panel (b)]. Noting
Eq.(\ref{eq:EktVAlim}), we can attribute this new accumulating point to
the E mode, particularly to the constant residual E mode. That is,
two $\omega=0$ solutions exist: one from the continuum CvK mode and
another from the new residual E mode, which causes the
singularity at $\omega_r=0$ in the spectral density figure [panel
(b)]. Panels (c) and (d) show the corresponding eigen functions
$\delta f(v_j)$. As expected, the eigen function of the continuum
CvK mode is singular\cite{Kampen1955, Case1959}, whereas the eigen
function of the E mode is smoother [Due to the numerical error
($\omega\simeq10^{-16}\neq0$), we do not know whether the $\delta$-singularity at
$v=0$ in panel (c) is a correct structure or a numerical problem at
present. The values $\Delta^{+}(\delta f_{v_j=0})\equiv\delta f(\Delta v)-\delta f(0)$
or $\Delta^{-}(\delta f_{v_j=0})\equiv\delta f(0)-\delta f(-\Delta v)$ of the peaks in
panels (c) and (d) are also sensitive to the discrete grid size $\Delta v$.].

\section{Summary}\label{sec:summary}

In this paper, we have presented a relatively complete picture for
the modes in the linearized electrostatic 1D V-A system. Besides the
usual Landau mode and the ballistic mode in the V-P system, a new
constant residual mode is found, and this mode is usually dominant.
Analytical asymptotic solutions, eigenmode solutions, and linear
simulations are consistent. In contrast to ballistic mode, this
residual mode is more robust, i.e., more difficult to be damped.

The finding indicates that the residual mode would be common, and not
merely exists in complicated system such as the Rosenbluth-Hinton residual
zonal flow found in toroidal system.

One may also be interested in other applications of V-A system. A
successful example of the application of extended V-A equation with
collision, source, and sink is the Berk-Breizman
model\cite{Berk1990,Berk1996} for discussing the nonlinear single
Alfv\'en mode driven by an energetic injected beam, especially for
Alfv\'en eigenmodes in tokamak.

\section{Acknowledgements}
The author would like to thank M. Y. Yu for his assistance in
improving the manuscript and for the helpful discussions.
Discussions with R. B. Zhang and comments from Y. Xiao are also
appreciated. We especially thank the two anonymous referees for
their useful comments and suggestions, which are helpful in
improving this work. This work is supported by Fundamental Research
Fund for Chinese Central Universities.

\appendix

\section{Vlasov-Poisson system}\label{sec:vp}
For Landau mode, the dispersion relation of the V-P system
Eq.(\ref{eq:es1d}) is
\begin{equation}\label{eq:drvp}
    D_{\mbox{VP}}(\omega,k) \equiv k+\int_C{\frac{{\partial
    f_0}/{\partial v}}{\omega-kv}dv}=0,
\end{equation}
where $C$ is the proper integral
contour\cite{Landau1946,Chen1987,Jackson1960,Krall1973,Nicholson1983}.

Similar to Chen\cite{Chen1987} or Jackson\cite{Jackson1960},
by applying Laplace transform ($L_p$) in time and Fourier transform
($F_r$) in space to the V-P system, we can obtain
\begin{equation}\label{eq:EktVP}
    \delta E(t,k) = \int_{C_\omega}{\frac{e^{-i\omega t}d\omega}{2\pi}}\Big [\int{dv\frac{\delta f_k(0)}{(\omega-kv)D(\omega,k)}}\Big].
\end{equation}

We are interested in asymptotic behavior, so that only the
$\omega=kv$ pole and the maximum-$\Im\omega_m$ normal mode are kept.
From Eq.(\ref{eq:EktVP}) we have
\begin{equation}\label{eq:EktVPlim}
    \lim_{t \to \infty}\delta E = {\frac{1}{2\pi}}\int_{v_L}{dv\Big [\underbrace{\frac{e^{-ikvt}\delta f_k(0)}{D_{\mbox{VP}}(kv,k)}}_{\mbox{ballistic part}} +\underbrace{\frac{e^{-i\omega_mt}}{(\omega_m-kv)\partial_{\omega_m} D}}_{\mbox{Landau
    part}}
    \Big]}.
\end{equation}

Focusing on the ballistic part, one can easily solve for the
Gaussian perturbation or Eq.(\ref{eq:df0}), and obatain the
$\propto\exp(-t^2)$ or $\propto1/t$ solutions.

\begin{figure}
  \includegraphics[width=8.5cm]{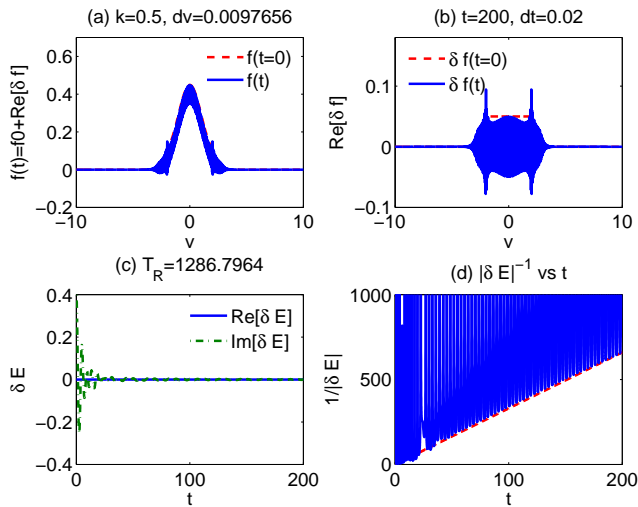}\\
  \caption{Simulation of the $1/t$ decay ballistic mode. (a) and (b) show
  the $f(v,t)$ and $\delta f(v,t)$ at $t=0$ and $t=200$. (c) and (d) show
  the $\delta E(t)$ and $|\delta E(t)|^{-1}$ versus $t$. The $\propto 1/t$ behavior
  of $\delta E(t)$ is clear in (d), where the red dashed line is $y=3.3x$ and is provided as reference.}\label{fig:vp_ballistic}
\end{figure}

The analytical solutions from Eqs.(\ref{eq:EktVP}) or
(\ref{eq:EktVPlim}) are usually approximations, so we would like to
verify them numerically.

We solve Eq.(\ref{eq:es1d}) as an initial value problem from $t=0$ to
$t_{\rm end}=N_t\Delta{t}$ using a 4th-order Runge-Kutta scheme. The
discrete velocity space is from $v_{\min}$ to $v_{\max}$. There are
$N_v$ uniform grids of size $\Delta{v}=(v_{\max}-v_{\min})/N_v$.

As mentioned, for most initial perturbations $\delta{f}(t=0)$, the
asymptotic behavior of $\delta{E}(t)$ is determined by the normal
mode that is Landau damped. Using a Gaussian initial perturbation
$\delta f$ with $A_0=0.05$, $u_a=1.0$, and $u_b=1.0$, we have
successfully reproduced the Landau damped solution (not shown here,
or see Fig.\ \ref{fig:va_landau}). The electric field $\delta
E(t=0)$ and $\delta E(t)$ are obtained from the Poisson equation
using the initially given and the calculated $\delta f$,
respectively. We should use a small $\Delta v$ to avoid
non-physical recurrence effect (the Poincar\'e recurrence) at
$T_R=2\pi/(k\Delta v)$, which is due to the discreteness of the
velocity space \cite{Cheng1976}. The verification also indicates that
our simulation scheme is feasible.

The $1/t$ damped ballistic mode from the initial perturbation
Eq.(\ref{eq:df0}) can be seen in Fig.\ \ref{fig:vp_ballistic} for
$A_0=0.05$, $u_a=0.0$, and $u_b=2.0$. As reference, the red dashed
line in panel (d) is for $y=3.3x$. $|\delta E|$
decays as $c/t$, as analytically predicted in Eq.(\ref{eq:timeasym}).
Note that the constant $c$ depends on $A_0$.

Notably, our linear analysis and simulation are
carried out in the $k$-space, i.e., all perturbation quantities have
an $e^{ikx}$ factor. However, one can obtain the corresponding $(x,v)$
phase space figure by direct mapping using the relation
$f(x,v)=f_0(v)+\Re[\delta f(k,v)e^{ikx}]$.

The above analytical calculations and simulations show a
complete picture of the typical Landau and ballistic modes in the V-P
system.

\end{CJK*}

\end{document}